\documentstyle[aps,12pt,epsfig,epsf,psfig19,floats]{revtex}

\newcommand{\ageq}{\mbox{\
\raisebox{-.9ex}{$\stackrel{\textstyle >}{\sim}$}\ }}

\def\om{{\omega}}

\def\begineq{\begin{equation}}
\def\be{\begin{equation}}
\def\endeq{\end{equation}}
\def\ee{\end{equation}}

\begin{document}
\pagenumbering{arabic}

\bibliographystyle{prsty}

\title{Response of bubbles to diagnostic ultrasound:\\ a unifying
theoretical approach}

\author{
Sascha\,Hilgenfeldt and Detlef\,Lohse }
\address{
Fachbereich Physik der Universit\"at Marburg,
Renthof 6, 35032 Marburg, Germany}
\author{Michael Zomack}
\address{Schering AG, Clinical Development, M\"ullerstr.\ 178,
13342 Berlin, Germany}

\maketitle

\vspace{0.5cm}

\begin{abstract}
The scattering of ultrasound from bubbles of $\sim 1\mu m$ radius, such
as used in contrast enhancers for ultrasound diagnostics, is studied.
We show that sound scattering and ``active'' emission of sound from
oscillating bubbles are not
contradictory, but are just two different aspects derived from the same
physics. Treating the bubble as a nonlinear oscillator, we arrive at
general formulas for scattering and absorption cross sections. We show that
several well-known formulas are recovered in the linear limit of this
ansatz. In the case of
strongly nonlinear oscillations, however,
the cross sections can be larger than those for linear response by several
orders of magnitude. The major part
of the incident sound energy is then converted into emitted sound, unlike
what happens in the linear case, where the absorption cross sections exceed
the scattering cross sections.

\vspace{0.5cm}


\end{abstract}

\section{Introduction}
In recent years, a new kind of contrast agent for use
in ultrasound diagnostics has been developed: suspensions of gas filled
microbubbles, i.e., bubbles of at most a few micrometers in diameter
\cite{nan93}.
The enhancement of image brightness and contrast is mostly due to the
well-known fact that microbubbles of this size can have
tremendous scattering cross sections for the incident diagnostic ultrasound
(with frequencies around $1-10$MHz) \cite{gra68}.
A number of different theoretical approaches are present in literature,
resulting in
formulas for the scattering and absorption cross sections that do not
always seem
compatible (see e.g.\ \cite{nan93,ll87,lei96,nis75}). It is one of the
main goals of this work to show that
all of these results can be understood as special cases of
a unified approach which treats the bubble as a (generally nonlinear)
volume oscillator. The different appearance in literature is only 
due to the use of different formalisms (e.g.\ full-fledged scattering theory
with partial wave decomposition \cite{nis75} or linear oscillator theory
\cite{lei96})
and the validity of different
limiting cases (e.g.\ neglecting surface tension \cite{nan93,ll87}).

The other main goal of our present study
is to go beyond previous work \cite{gro97_zom}, where monochromatic
driving was treated, and employ
{\em pulsed} driving within our formalism, in order to achieve a more realistic
modeling of the situation in diagnostic applications. The analysis of the
emitted sound from the bubbles in terms of intensity and spectral distribution
is of obvious importance to assess the
signal quality of ultrasonography with bubble contrast agents. While a
large number of publications have dealt with the sound emission of
{\em linearly} oscillating bubbles (see e.g.\ \cite{lei96,nis75} and references
therein), and the {\em weakly} nonlinear case was treated by
Prosperetti \cite{pro74,pro75}, 
a systematic study of cross sections for {\em strongly nonlinear} response to
{\em pulsed} driving has, to our knowledge, not yet been performed.
It is this fully nonlinear case of bubble oscillations which is encountered
in ultrasound diagnostics as driving pressure
amplitudes of up to $\sim 10$\,atm
are common, whereas strong nonlinearities occur already for amplitudes
$\ageq 1$\,atm.

We present our general, unifying approach in Section~\ref{soundemabs} and
show in Section~\ref{lincase} how it translates to a multitude of linear
limiting cases, where analytical results can be obtained if
monochromatic driving is assumed.
Still in the linear limit, we then introduce pulsed driving in \ref{pulses}
and finally present results for the full nonlinear case in Section~\ref{fullnl}.
A more detailed discussion, especially of spectral properties of the emitted
sound for nonlinear response, will be postponed to another publication
\cite{hil98}. Section~\ref{summary} presents conclusions.

\section{Sound emission and absorption}\label{soundemabs}

\subsection{Emitted sound pressure}

We want to evaluate the pressure of emitted sound $P_s(r,t)$
from a body capable of volume oscillations (i.e., a bubble) driven by an
incident pressure wave $P(t)$.
First, we notice that there are two contributions to
$P_s(r,t)$: (i) the {\em active emission} of sound, caused by the
change of volume of the body, and (ii) the {\em passive} contribution
due to the mere presence of the (maybe non-oscillating) body in the
incident field. Thus, we can write
\begineq
P_s(r,t)= P_s^a(r,t)+P_s^p(r,t)\,.
\label{ps}
\endeq

For a body of volume $V_b(t)$ which is
much smaller than the wavelength of the incident sound, the pressure
$P_s^a(r,t)$ of {\em actively
emitted} sound is given to leading order by \cite{ll87,bre95b}
\begineq
P_s^a(r,t)={\rho_l \over 4\pi r} {d^2V_b \over dt^2}\,,
\label{vbt}
\endeq
Here, $\rho_l$ is the liquid density  and we have adopted a spherical
coordinate system with radius $r$. There is no angular dependence because
we have assumed the long wavelength limit (S wave scattering).
For a spherical bubble with radius dynamics $R(t)$,
(\ref{vbt}) easily translates to
\begineq
P_s^a(r,t)=\rho_l {R(t)\over r}
\left( 2 \dot R(t)^2 + R(t)\ddot{R}(t)\right)={1\over r}q_s^a(t)\,,
\label{psa}
\endeq
where we have introduced $q_s^a(t)\equiv rP_s^a(r,t)$, thus separating
the trivial (geometrical) $1/r$ spatial dependence of the radiation
from the time-dependent part.

The second term $P_s^p(r,t)$ in (\ref{ps}) is the pressure of
{\em passively} emitted sound due to the mere presence of the body. It arises
from the perturbation of the density field in water: if the body was not
present, the incident sound wave could induce density changes in the volume of
liquid occupied by the body. As this cannot happen, the situation is
equivalent to having an oscillator that compensates the effects of the
sound wave; this oscillator has a (virtual) volume change
\be
{dV\over dt}=V_0{d\rho_l/dt\over \rho_l}\,,
\label{virtvol}
\ee
with the (now constant) volume $V_0$ of the body, see \cite{ll87}.
For a sound wave $P(t)$, we have $dP/d\rho_l=c_l^2$, where the
derivative is taken at constant entropy. Then, in analogy to (\ref{vbt}),
replacing $V_b$ by $V$ and inserting $V_0=4\pi R^3/3$, we get the
leading order
pressure change due to the passive reaction to the sound wave 
\begineq
P_s^p(r,t)={1\over 3rc_l^2}R^3(t)\ddot{P}(t)={1\over r}q_s^p(t)\,,
\label{psp}
\endeq
where $c_l$ is the speed of sound in the liquid. Note again that we
are in the limit $R(t)\ll\lambda$, where  $\lambda$ is the wavelength
of incident sound. Therefore, the driving pressure $P(t)$ can be
treated as spatially uniform, i.e., the pressure experienced by the bubble
does not vary over its size. In analogy to (\ref{psa}), we
have defined the quantity $q_s^p(t)$.

\subsection{Bubble oscillation}

To evaluate (\ref{psa}) or (\ref{psp}), we need a formula for $R(t)$.
The oscillatory behavior of the radius $R$ of a gas bubble in a liquid is well
described by the Rayleigh-Plesset equation \cite{ray17,bre95b}.
Many variants of
this equation have been presented
(see e.g.\ \cite{gil52,fly75,kel80,las81,las82,pro91,pro94});
the following form
has proved to be robust and close to experiment even in situations of
massively nonlinear bubble behavior, as e.g.\ in sonoluminescence
experiments \cite{bar97}:
\be
R \ddot R + {3\over 2} \dot R^2
=
{1\over \rho_l} \left(p(R,t) - P(t) - P_0 \right)
+
		    {R\over \rho_l c_l} {d\over dt}
                    p(R,t) - 4 \nu_l{\dot R \over R}-{2\zeta \over\rho_l R}\,.
                    \label{rp}
\ee
We have introduced the liquid viscosity $\nu_l$ and surface tension $\zeta$
here, as well as the ambient pressure $P_0$ taken to be 1\,atm in this
work. All liquid parameters assume the values of water, except for
$\nu_l$, which is multiplied by three to mimic the viscosity of blood
\cite{lan69}.
$p(R,t)$ stands for the gas pressure {\em inside} the bubble and
is modeled by a polytropic process equation with van der Waals hard core
and parameters for air. The polytropic exponent is taken to be one, because
bubbles of the sizes we treat here ($\sim 1\mu m$) are smaller than the
thermal diffusion length on the time scales of the oscillation, and can
therefore be regarded as isothermal\cite{ple77,hil96} .

\subsection{Scattering cross sections}\label{scattcrosssec}

In general, the scattering cross section $\sigma_{sc}$ is related to
the incident intensity (energy/area/time) $I_{inc}$ and the scattered
power $W_{sc}$ (energy/time) via
\begineq
W_{sc}=\sigma_{sc}\,I_{inc}\,,
\label{sigma}
\endeq
and has thus the dimensions of an area.
$I_{inc}$ is determined by the incident pressure wave $P(t)$,
namely (see, e.g., \cite{ll87}, \S 65)
\begineq
I_{inc}={1\over\rho_l c_l}\langle P^2(t)\rangle_t
\label{iinc}
\endeq
and $W_{sc}$ follows from $P_s(r,t)$ through
\begineq
W_{sc}={4\pi\over\rho_l c_l}
 \langle r^2P_s^2(r,t)\rangle_t
 ={4\pi\over\rho_l c_l}\langle q_s^2(t)\rangle_t\,.
 \label{wsc}
\endeq
Here, $\langle\cdot\rangle_t$ denotes a time mean; we have exploited
the virial theorem to express intensities and powers solely in terms
of pressures (the ``potential'' part of the energy of the wave).
The integral
\begineq
E_{sc}= {4\pi\over \rho_l c_l}\int_0^\tau q_s^2(t)\,dt
\label{esc}
\endeq
gives the total scattered energy over a time span $\tau$ (e.g.\ the
duration of an incident sound pulse).

From these definitions, it is clear that the general formula for
the scattering cross section is
\begineq
\sigma_{sc}={4\pi\over\langle P^2(t)\rangle_t}
\langle q_s^2(t)\rangle_t=
{4\pi\over\langle P^2(t)\rangle_t}
\langle \left[ q_s^a(t)+q_s^p(t)\right]^2 \rangle_t\,.
\label{sigmasc}
\endeq
We notice that the active and passive parts {\em interfere}.
It is therefore, strictly speaking, wrong
to divide $\sigma_{sc}$ into
an active and a passive part, as it is sometimes done in literature. 
Nevertheless,
it will be shown below that in the case of diagnostic bubbles
the total scattering cross section is almost exclusively
due to the active part $P_s^a$ of (\ref{ps}), while the passive contribution
can safely be neglected.

There are other possible contributions to $P_s(r,t)$, e.g.\ direct or
indirect results of bubble shape oscillations. These are not analyzed
here because perfect sphericity is assumed. Also, there is a contribution
if the bubble (or the body in general) can be translated as a whole by
the incident sound. This term is orthogonal to those treated above
(i.e., the interference terms vanish) and results in
a well-defined {\em additional} scattering cross section (cf. \cite{ll87}
\S 78 or \cite{nan93}) with a characteristic angular dependence. As it is
(for the case of a gas bubble in a liquid) always of the same order of
magnitude as the passive contribution, and is therefore equally negligible,
we do not treat it here in detail.
Finally, if two or more bubbles come close to each other, they will
change the emitted sound field either by direct secondary scattering or
indirectly by modifying their modes of oscillation
(e.g.\ via secondary Bjerknes forces \cite{bje09,pro84b}).
We will not try to incorporate
these effects, but restrict ourselves to the case of a single
bubble.

\subsection{Absorption cross section}
We have so far regarded the scattering cross section as an indicator of
the acoustic energy deflected by the scatterer.
Likewise, the absorption cross
section $\sigma_{abs}^\nu$ stands for the energy loss induced by the
viscous term in the RP equation (\ref{rp}). In
this case the energy is directly converted into heat and will, in general,
be of little use to the experimenter, whereas the scattered
sound can be detected much more easily. Nevertheless, $\sigma_{abs}^{\nu}$ 
is important in
order to assess the energy balance of the scattering process.
The liquid viscosity
exerts a stress
\begineq
p_{vis}={4\nu_l\rho_l\dot{R}\over R}
\label{viscstress}
\endeq
over the bubble surface of size $A=4\pi R^2$ (this can be read off directly
from the RP equation). As the bubble wall moves
with velocity $\dot{R}$, the dissipated power is
\begineq
W_{dis}^{\nu}=p_{vis} A \dot{R} = 16\pi\nu_l\rho_l R\dot{R}^2\,.
\label{viscdiss}
\endeq
The absorption cross section is determined in analogy to $\sigma_{sc}$
via
\begineq
W_{dis}^{\nu}=\sigma_{abs}^{\nu}\,I_{inc}\,.
\label{sigmaabs}
\endeq

\section{Small driving: the linear case}\label{lincase}

The formulas given in the previous sections can be applied to all
cases of bubble motion and subsequent sound emission.
For small driving, they reduce to the linear case and almost all quantities
can be calculated analytically.
First of all, the RP equation can be linearized:
we set $R(t)=R_0(1+x(t))$ ($R_0$ is the radius of the undriven bubble
under ambient conditions) and get 
\begineq
\ddot{x}+2\gamma\dot{x}+\om_0^2 x={P(t)\over\rho_l R_0^2}
\label{rplin}
\endeq
with the viscous damping constant $\gamma={2\nu_l\over R_0^2}$ and the
bubble eigenfrequency
\begineq
\om_0^2={3\kappa_g P_0\over\rho_l R_0^2} + {4\zeta\over\rho_l R_0^3}\,.
\label{om0}
\endeq
This latter quantity consists of two distinct terms, the first due to
the gas pressure $p(R,t)$, and the second governed by surface tension $\zeta$.
For $p(R,t)$, a polytropic ideal gas formula was chosen. While it is
advisable to employ a more elaborate formula (e.g.\ a van der Waals gas
with hard core) if the bubble oscillation is violent, the ideal gas is a
very good approximation in the linear limit. The
polytropic exponent $\kappa_g$ measures the gas compressibility
($p(R,t)\propto\rho_g^{\kappa_g}$); it is 1 for isothermal behavior and
equal to the adiabatic exponent for adiabatic behavior of an ideal gas
(e.g.\ 7/5 for air). In deriving Eq.~(\ref{rplin}), we note that effects
of thermal damping are neglected (as they have not been present in (\ref{rp}),
either), and also radiation damping (caused in (\ref{rp}) by the term
involving $c_l$) is not present. Both of these damping effects
are very small in the parameter range
of our interest; thus, $\gamma$ only contains viscous damping contributions.

\begin{figure}[htb]
\setlength{\unitlength}{1.0cm}
\begin{center}
\begin{picture}(10,8)
\put(-1.,0.){\psfig{figure=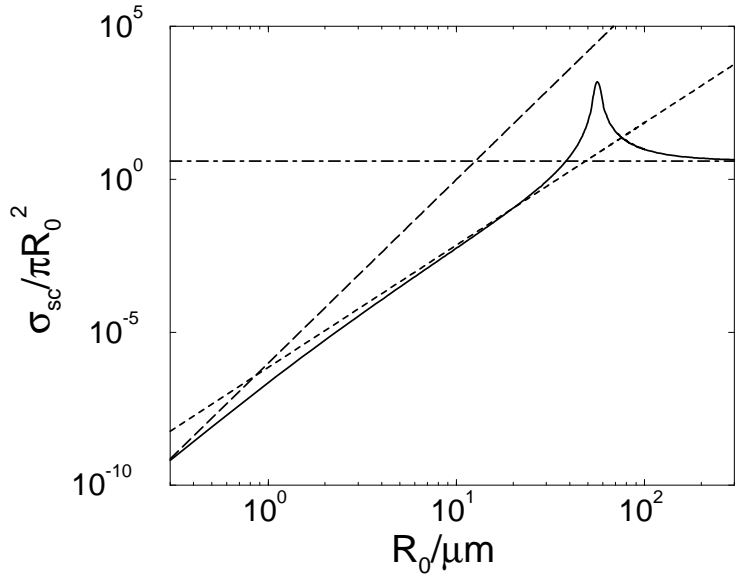,width=10cm,angle=-90}}
\end{picture}
\end{center}
\caption[]{Numerical computation (solid line)
and scaling behaviors of the normalized
scattering cross section for bubbles driven with a monochromatic $50$kHz
sound wave at small drive amplitudes (linear case).
We chose a representation similar to that of Nishi \cite{nis75},
displaying $\sigma_{sc}$ divided by the geometrical cross section
$\pi R_0^2$ on a double logarithmic plot.
To the right of the resonance peak (for large $R_0$),
$\sigma_{sc}/(\pi R_0^2)$ relaxes to 4 (dot-dashed line). In the
opposite limit of small $R_0$, it scales $\propto R_0^4$ (dashed)
for an interval,
but becomes steeper again for very small $R_0$, reaching an asymptote
$\propto R_0^6$ (long dashed).
These scaling laws correspond to the $R_0^6$ and $R_0^8$
behaviors of the unnormalized $\sigma_{sc}$, respectively
(cf.\ Eqs.~(\ref{smallrgas}),(\ref{smallrsurf})).}
\label{50khz}
\end{figure}

To get explicit solutions, let us assume monochromatic driving
$P(t)=P_a\cos \om_d t$ in (\ref{rplin}). $P_a$ is the driving pressure
amplitude, $\om_d=2\pi f_d$ the angular driving frequency. This yields
\begineq
x(t)=\varepsilon \cos(\om_d t + \delta)\\
\label{xsol}
\endeq
with the amplitude
\begineq
\varepsilon=-{P_a\over\rho_l R_0^2}
{1\over\sqrt{(\om_0^2-\om_d^2)^2+4\gamma^2\om_d^2}},
\label{xamp}
\endeq
and the phase shift
\begineq
\delta=\arctan\left({2\gamma\om_d\over\om_d^2-\om_0^2}\right)\,.
\label{xphase}
\endeq
Obviously, in the monochromatic case $I_{inc}={P_a^2\over 2\rho_l c_l}$.
With $R(t)$ from $x(t)$, (\ref{psa}), (\ref{psp}), and (\ref{sigmasc})
we get
\begineq
\sigma_{sc}^{lin}=4\pi R_0^2 \left<\left[
{\sqrt{2}\om_d^2\cos(\om_dt+\delta)\over
\sqrt{(\om_0^2-\om_d^2)^2+4\gamma^2\om_d^2}}- {\sqrt{2}\over3c_l^2}
R_0^2\om_d^2\cos\om_d t\right]^2\right>_t
\,.
\label{lincrossfull}
\endeq
Note that
$\sigma_{sc}^{lin}$ does not depend on the absolute size of $P_a$, as
long as the oscillation stays linear.
Fortunately, in all cases of interest to bubble ultrasound diagnostics,
i.e., for frequencies in the MHz range and $R_0$ between a few tenths
of a micron and a few microns,
the second (passive) term of this formula is negligible compared to the
first (active) part. We will show in detail below that the dominance of
the active emission is so overwhelming that the relative
errors in neglecting the passive contribution never exceed $0.25\%$ in the
parameter range of this study. Dropping the
passive term, we get the considerably simplified expression
(reproducing results presented e.g.\ in Leighton \cite{lei96} in Section 4.1)
\begineq
\sigma_{sc}^{lin}\approx
4\pi R_0^2 {\om_d^4\over (\om_0^2-\om_d^2)^2+4\gamma^2\om_d^2}\,.
\label{lincross}
\endeq
This formula can also be translated into the S wave scattering limit
of the complete scattering analysis by Nishi \cite{nis75}. In equation~(47) of
that work, there are additional contributions to $\gamma$ due to thermal
and radiation damping, as well as modifications involving the stiffness
of the bubble.
In certain limiting cases, (\ref{lincross}) gets further simplified. Moreover,
a rich variety of scaling behaviors, especially in $R_0$, can be found.
The richness (as compared to ``ordinary'' linear oscillators) is due to
the fact that in (\ref{rplin}), the damping, driving frequency and
dimensionless driving {\em all} depend on $R_0$. In Fig.~\ref{50khz}
$\sigma_{sc}^{lin}$ is presented for a $50$kHz-driven bubble, a frequency
chosen such that most of the limiting cases treated in the following
subsections can be illustrated by the graph. Note that all cross section
graphs in this work were computed with the full numerical formulas
presented in Section~\ref{soundemabs}.

\subsection{Small bubbles: dominance of gas pressure or surface tension}

For small bubbles (such that $\om_0\gg\om_d$) the denominator of
the active contribution in (\ref{lincrossfull}) simplifies to $\om_0^2$.
It is crucial here to verify which of the two
contributions to $\om_0$ in (\ref{om0}) is dominant. If 
surface tension is very small (and $R_0$ is not extremely tiny),
$\om_0$ will be governed by the first term, i.e., the one due to the
internal gas pressure. 
With the polytropic ansatz, we can
use $c_g^2=\kappa_g P_0/\rho_g$ for the speed of sound in the gas $c_g$ and
from (\ref{lincrossfull}) the limiting behavior for small $R_0$ is then
\begineq
\sigma_{sc}^{lin}\to {4\pi\over 9} {R_0^6\om_d^4\over c_l^4}
\left(1-{K_g\over K_l}\right)^2
\label{smallr}
\endeq
with the compressibilities $K_i=1/(\rho_ic_i^2)$ for the bubble interior
($i=g$) and exterior ($i=l$). This reproduces the
equations presented e.g.\ by Landau \cite{ll87} \S 78, deJong \cite{dej91} or
Meerbaum in Nanda/Schlief \cite{nan93} (apart from the translational
contribution mentioned in \ref{scattcrosssec} above).
Note that in the case of gas bubbles in a liquid, $K_g$ will be much greater
than $K_l$, so that to very good approximation
\begineq
\sigma_{sc}^{lin}\to
{4\pi\over 9} {R_0^6\om_d^4\over c_l^4}\left({K_g\over K_l}\right)^2
\label{smallrgas}
\endeq
and the dominance of the active scattering is again confirmed. In fact,
as for typical materials $\rho_l\approx 1000\rho_g$ and $c_l\approx 5 c_g$,
the active emission can surmount what would be expected from a purely
passive scatterer by more than eight orders of magnitude! This is why
oscillating bubbles are so much superior to (completely passive) hard
spheres in terms of ultrasound scattering capability.
In Fig.~\ref{50khz} there is indeed a region where the $\propto R_0^6$
behavior of $\sigma_{sc}^{lin}$ is observed.
Note, however, that the validity of (\ref{smallr}) and (\ref{smallrgas}) for
bubbles is limited, especially as surface tension is explicitly neglected.

If surface tension is taken into account and the parameters for water
are inserted into (\ref{om0}), we see that for
$R_0\leq 0.96\mu m$ the
surface tension term dominates $\om_0$, so that 
$\om_0^2\to 4\zeta/\rho_l R_0^3$. As a consequence, the active
scattering cross section will acquire a different limit, namely
\begineq
\sigma_{sc}^{lin}\to {\pi R_0^8\om_d^4\rho_l^2\over 4\zeta^2}\,.
\label{smallrsurf}
\endeq
The steeper $\propto R_0^8$ behavior takes over in Fig.~\ref{50khz} for
very small bubbles, as expected.
As the passive contributions scale like
$R_0^6$, there has to be a (small) radius where the latter become dominant.
This radius is easily calculated to be
\begineq
R_0^p={4\zeta \over 3\rho_l c_l^2}\,,
\label{crosspass}
\endeq
which, for the material constants of the water/air (blood/air) system, is
$\approx 4.4\cdot 10^{-11}m$, i.e., on a subatomic scale which is not
described by the physics discussed here.

\subsection{Linear resonance}
When $R_0$ is such that $\om_0\sim\om_d$, the bubble is near resonance,
i.e., its oscillation amplitude $\epsilon$ and thus $\sigma_{sc}^{lin}$
become
much larger than for neighboring $R_0$. E.g., for the case of air bubbles in
blood driven with $\om_d=2\pi\cdot 3$\,MHz, it reaches values of about
15 times the geometrical cross section
$\pi R_0^2$ (cf.\ Fig.~\ref{lin3mhz} below),
i.e., the normalized cross section $\sigma_{sc}^{lin}/(\pi R_0^2)$
is larger than for any other radius. From (\ref{lincross}), 
it is easy to get
\begineq
\sigma_{sc}^{lin,res}\approx
 {\pi (R_0^{res})^6\om_d^2\over 4 \nu_l^2}\,,
\label{scattlinres}
\endeq
where $R_0^{res}$ is determined by the condition $\om_0(R_0^{res})=\om_d$.
The height of the resonance 
peak is thus mainly determined by the strength of damping, i.e., the viscosity
of the liquid. In water with its three times lower viscosity, the maximum
cross section is therefore 9 times higher than in blood.
The (Lorentz shaped) resonance peak is
the hallmark of the classical pictures of linear bubble scattering cross
sections (see \cite{nan93,lei96,nis75}). 
We will see that this shape changes considerably in the nonlinear
case.

\subsection{Large bubbles: the ``soft sphere'' limit}

For large bubbles (but still obeying $R\ll\lambda$), $\om_d\gg\om_0$ and
\begineq
\sigma_{sc}^{lin}\rightarrow 4\pi R_0^2,
\label{sigmasoft}
\endeq
i.e., the scattering cross section becomes four times the geometrical
cross section, which is also reproduced in Fig.~\ref{50khz}.
Equation~(\ref{sigmasoft}) describes the behavior for {\em all} $R_0$
if the bubble
interior is arbitrarily {\em compressible} while surface tension is absent
($\zeta=0$).
This can be seen if we imagine $\kappa_g\to 0$ (implying $c_g\to 0$ and
$K_g\to\infty$)
and thus $\om_0\to 0$:
the resonance peak is shifted to very small $R_0$ and every larger
ambient radius
results in the ``soft sphere limit'' cross section (\ref{sigmasoft}). 

In reality, for large bubbles there are of course still the interference
contributions with
the passive term $\propto R_0^6$. But the influence of $P_s^p$ can only
outweigh the active emission for $R_0>{\sqrt{3}\over 2\pi}\lambda$ (which
is again easy to prove);
this is already in a region where the assumption $R_0\ll\lambda$ is dubious.
In fact, for diagnostic ultrasound with $\om_d=2\pi\cdot 3$MHz,
$\lambda\approx 500\mu m$, resulting in a critical radius for passive
influence of $R_0\ageq 140\mu m$.
Thus, there is at most a very small region of transition until the scattering
cross section has to be described by completely different formulas (Mie
scattering theory), and the
bubbles in this region are of no interest to ultrasound diagnostics because
of their size.

\subsection{Other limits}

There are a number of limiting cases of (\ref{lincrossfull}) and
(\ref{lincross}) that go beyond the parameter space of this study; they are 
given here for completeness:
\begin{itemize}
\item
In the limit of $R_0\gg\lambda$, the cross section relaxes to the
limit $\sigma_{sc}=2\pi R_0^2$ \cite{nis75}. This is only relevant for
mm-sized or even larger bubbles.
\item
If we apply our formulas to solid spheres in a liquid rather than bubbles,
i.e., we make the interior of the ``bubble'' much more stiff than
the liquid, the main contribution in (\ref{smallr}) is the {\em passive}
scattering, as $K_g/K_l\ll 1$.
The resulting cross section is the ``hard sphere'' Rayleigh limit
\begineq
\sigma_{sc}^{Ra}={4\pi\over 9} {R_0^6\om_d^4\over c_l^4}\,.
\label{hardray}
\endeq
In analogy to the case of the ``soft sphere'' above, the resonance peak
is shifted to arbitrarily {\em large} radii for an arbitrarily stiff
bubble, so that here (\ref{hardray}) is valid for every finite $R_0$. 
\end{itemize}

\subsection{The regime of ultrasound diagnostics}

\begin{figure}[htb]
\setlength{\unitlength}{1.0cm}
\begin{center}
\begin{picture}(10,8)
\put(-1.,0.){\psfig{figure=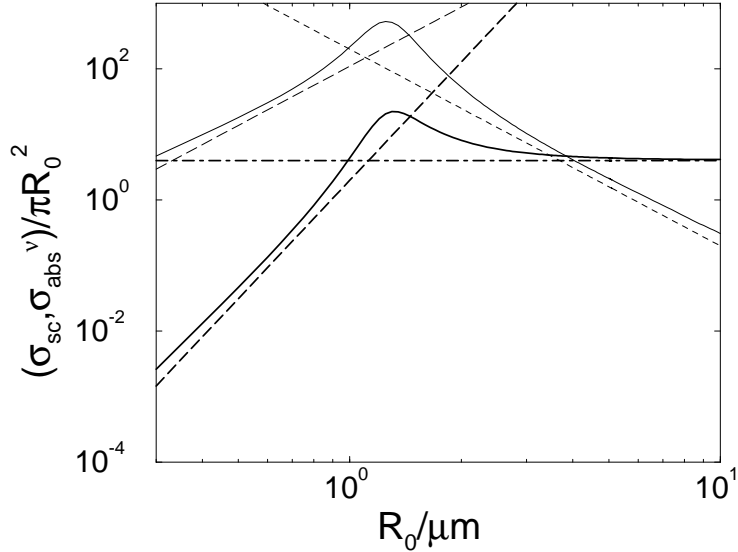,width=10cm,angle=-90.}}
\end{picture}
\end{center}
\caption[]{Numerical computation (solid lines)
and scaling behaviors of the normalized linear
scattering (thick lines) and absorption (thin lines)
cross sections for bubbles driven
with a monochromatic $3$MHz sound wave (diagnostic frequency).
The resonance features as well as the large $R_0$ limit of 4 (dot-dashed)
are again present in $\sigma_{sc}/(\pi R_0^2)$.
Of the small $R_0$ limits, only the surface tension governed
$\sigma_{sc}^{lin}\propto R_0^8$ remains (long dashed).
The corresponding scaling laws
for $\sigma_{abs}^{\nu,lin}/(\pi R_0^2)$ are $\propto R_0^{-1}$ (dashed)
for large and $\propto R_0^5$ (long dashed) for small $R_0$.
}
\label{lin3mhz}
\end{figure}

It is important to note that for the typical $\mu m$ size bubbles of
ultrasound diagnostics applications, the relevant limiting cases of
the linear formulas (\ref{lincrossfull}),\,(\ref{lincross}) are
(\ref{smallrsurf}),\,(\ref{scattlinres}), and (\ref{sigmasoft}). The
case of $\sigma_{sc}^{lin}\propto R_0^6$ given by (\ref{smallr}) does
not occur, because the resonance radii $R_0^{res}$ are -- for MHz driven
bubbles -- in the range of the crossover ambient radius $R_0\approx 0.96\mu m$
between gas pressure and surface tension dominated resonances. Thus, the
oscillations of bubbles
with $R_0<R_0^{res}$ are all surface tension dominated, and
never show the $R_0^6$ behavior. This is obvious in Fig.~\ref{lin3mhz}, which
shows the linear cross sections for 3MHz driven bubbles in blood. In contrast
to Fig.~\ref{50khz}, there is a direct transition from the resonance peak to
a curve $\sigma_{sc}^{lin} \propto R_0^8$.

\subsection{Linear absorption cross sections}

The linear analysis is naturally extended to absorption cross sections;
the analytical formula for
$\sigma_{abs}^{\nu}$ in the case of monochromatic driving is easily found
via (\ref{sigmaabs}) and (\ref{xsol})--(\ref{xphase}) to be
\begineq
\sigma_{abs}^{\nu,lin}={16\pi\nu_l c_l\over R_0}{\om_d^2\over
(\om_0^2-\om_d^2)^2+4\gamma^2\om_d^2}\,.
\label{abslin}
\endeq
Here, the limiting cases for small bubbles show
$\sigma_{abs}^{\nu,lin}\propto R_0^3$
for small $R_0$ for gas pressure dominated $\om_0$ and
$\sigma_{abs}^{\nu,lin}\propto R_0^5$
for the surface tension dominated case (see Fig.~\ref{lin3mhz}.
Thus, $\sigma_{abs}^{\nu,lin}\gg
\sigma_{sc}^{lin}$ for small bubbles, i.e., most of the energy incident
on the bubble is converted into heat through viscous damping, whereas only
a small part is available for sound emission. The already surprisingly high
{\em scattering} cross section for microbubbles is outnumbered considerably
by the {\em absorption} cross section.
The same is true in resonance,
where
\begineq
{\sigma_{abs}^{\nu,lin,res}\over\sigma_{sc}^{lin,res}}
\approx {4\nu_l c_l\over (R_0^{res})^3\om_d^2}\approx
23
\label{sigratiores}
\endeq
can be calculated for $f_d=3$MHz and the material constants of blood,
in very good agreement with numerical computation (cf.\ Fig.~\ref{lin3mhz}).

For large $R_0$ above the resonance radius we have
$\sigma_{abs}^{\nu,lin}\propto R_0^{-1}$; as $\sigma_{sc}^{lin}$ grows like
$R_0^2$, we have in this range
$\sigma_{abs}^{\nu,lin}\ll \sigma_{sc}^{lin}$, and the major part of the
energy goes into sound emission. 

\section{Pulsed driving}\label{pulses}

\subsection{The incident pulse}
One assumption that allowed us to give simple analytical formulas was
the monochromaticity of the driving. In reality, bubbles in ultrasound
diagnostics are driven by the signal of a diagnostic transducer, which
is almost always {\em pulsed}, and quite often (e.g.\ in
pulse wave doppler mode) the pulses are only a few wavelengths long.

\begin{figure}[htb]
\setlength{\unitlength}{1.0cm}
\begin{center}
\begin{picture}(10,10)
\put(-0.,-0.8){\psfig{figure=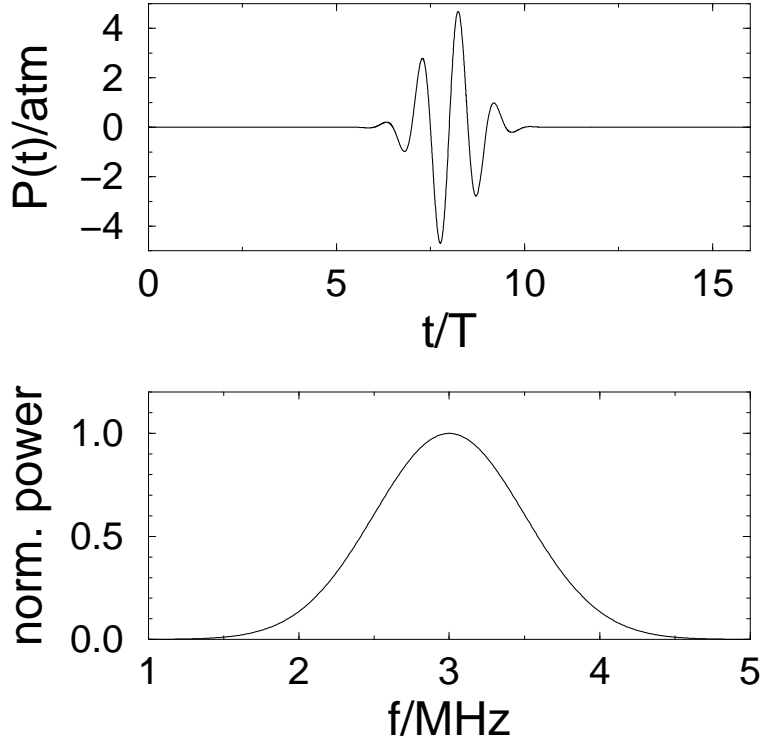,width=10cm,angle=0}}
\end{picture}
\end{center}
\caption[]{Modeled ultrasound diagnostics driving pulse. The upper part
of the figure shows the time series $P(t)$ ($t$ normalized to $T\equiv1/f_d$),
the lower part the normalized Fourier spectrum of the acoustic power given
by $P(t)$. It is centered around the mean frequency $f_d=3$\,MHz and decays to
$1/e^2$ of its maximum value within $hf_d=f_d/3=1$\,MHz above and below
$f_d$. This corresponds to the effective spatial extent of the pulse of
$\approx 3$ wavelengths. 
}
\label{figpulse}
\end{figure}

In closer agreement with
experimental reality, we therefore model the pressure pulse $P(t)$ as 
\begineq
P(t)=P_a\cos(\om_d (t-t_c))\exp\left(-{h^2\om_d^2\over 4} (t-t_c)^2\right)\,,
\label{pulse}
\endeq
centered around $t_c$ with relative width $h$. We choose $h=1/3$ here. After
Fourier transform, the frequency space representation is
\begineq
P(\om)={P_a\over h\om_d}\left[
\exp\left(-{(\om-\om_d)^2\over h^2\om_d^2}\right)
+\exp\left(-{(\om+\om_d)^2\over h^2\om_d^2}\right)\right]\,.
\label{pulsefour}
\endeq
This spectrum is (almost) Gaussian in shape; the corresponding power
$\propto |P(\om)|^2$ decays to $1/e^2$ of its maximum value within
a distance of $\pm h\om_d$ around $\om_d$ (cf.\ Fig.~\ref{figpulse}).
Thus, the reciprocal $1/h$ of the relative width is an approximate measure
for the spatial extension of the pulse (in wavelengths). In our example, the
pulse is about 3 wavelengths long and corresponds to the shortest pulses
routinely available in medical applications of diagnostic ultrasound.

\subsection{Effects on the cross sections}
As we are still in the linear limit, the pulsed driving can be understood
as a superposition of monochromatic waves of frequency $\om$
with amplitudes proportional to the Fourier components $P(\om)$ in
(\ref{pulsefour}). The response will be therefore obtained from a
convolution of the monochromatic response discussed above and the spectrum
$P(\om)$.

\begin{figure}[htb]
\setlength{\unitlength}{1.0cm}
\begin{center}
\begin{picture}(10,8)
\put(-1.,0.){\psfig{figure=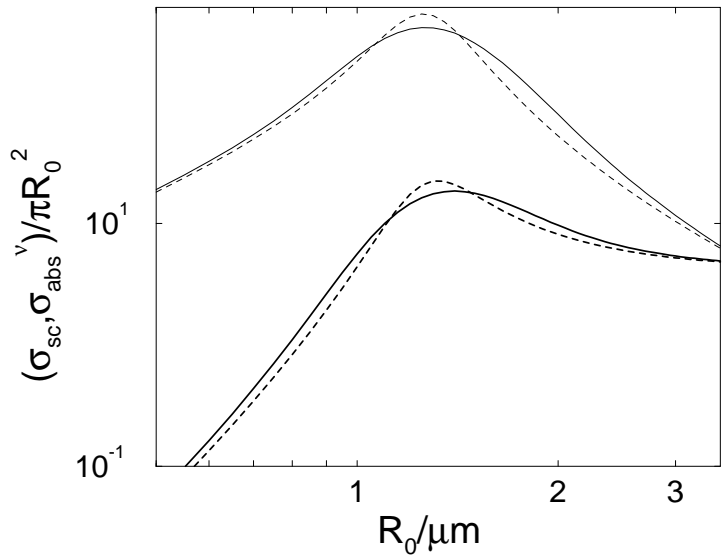,width=10cm,angle=-90.}}
\end{picture}
\end{center}
\caption[]{
Scattering (thick lines) and absorption (thin lines)
cross sections for bubbles driven
by a monochromatic 3\,MHz sound wave (dashed) and a diagnostic pulse
according to (\ref{pulse}) with center frequency 3\,MHz (solid).
The response to the polychromatic driving displays a broader, lower
resonance peak, which is shifted towards larger $R_0$. Further away from
resonance, the differences to the monochromatic case soon become
negligible.
}
\label{figmonopulse}
\end{figure}

Figure~\ref{figmonopulse} shows the cross sections
$\sigma_{sc}(R_0)$ and $\sigma_{abs}^{\nu}(R_0)$ for monochromatic and
pulsed driving in comparison. The figure is a
blow-up of the region near $R_0^{res}$, which is the only area where there
are marked differences between the responses to the different drivings.
As it may have been expected, the response curves for polychromatic driving
are broader, because unlike the monochromatic case, where the single
frequency $\om_d$ corresponds to a single, well-defined $R_0^{res}$,
there are a number of $R_0$ for which the bubbles react strongly to
the most intense frequency components in $P(\om)$. 
Correspondingly, a bubble driven at resonance by a monochromatic wave 
is a more effective scatterer than for pulsed driving, so that 
the maximum height of the cross section curves is smaller in the latter case.

Moreover, with pulsed driving there is a slight, but significant
shift in the maximum of both cross sections towards larger $R_0$. This
is due to the asymmetry in the response curve around the resonance
radius that is apparent already in the monochromatic case (cf.\
Fig.~\ref{figmonopulse}): on increasing $R_0$ from $R_0^{res}$, the
cross sections do not drop as rapidly as when $R_0$ is decreased from
the maximum of the curves. Therefore, the contributions in $P(\om)$ with
$\om$ slightly smaller than $\om_d$ (which correspond to
$R_0(\om)>R_0^{res}(\om_d)$) will have a larger effect on $\sigma_{sc}$ and
$\sigma_{abs}^{\nu}$ than the contributions with $\om$ slightly smaller
than $\om_d$, even if they are represented with the same weight in $P(\om)$.
Consequently, upon convolution of the spectrum, the cross sections are
increased for the larger $R_0$ and the maximum of the curves is shifted
towards $R_0>R_0^{res}(\om_d)$.

If the width of the pulse is varied, the peak shift and 
peak broadening vary correspondingly: thus, if the pulse becomes very short
(shorter than one wavelength), the resonance structure is almost completely
lost. Using longer and longer wavetrains (smaller $h$), on the other hand,
leads to successively sharper resonance peaks, while the repetitive 
oscillations of the bubble allow for the occurrence of subharmonic
resonances, provided that the bubbles remain shape stable 
(cf.~\cite{gro97_zom}).

\section{Results of full nonlinear computations}\label{fullnl}

Let us now come back to the general formulas presented in 
Section~\ref{soundemabs} and drop the assumption of linear response.
We will find that the features of the nonlinear case can
be quite different from the linear results presented above.
Not surprisingly,
the cross sections now depend on the driving pressure amplitude $P_a$,
which is not the case for linear driving.
It should be noted that, in order to compare these results with experimental
measurements, they have to be convoluted with the bubble size distribution,
as in virtually all experiments the bubble suspensions are not monodisperse.

\subsection{Scattering cross sections}
\begin{figure}[htb]
\setlength{\unitlength}{1.0cm}
\begin{center}
\begin{picture}(10,8)
\put(-1.,0.){\psfig{figure=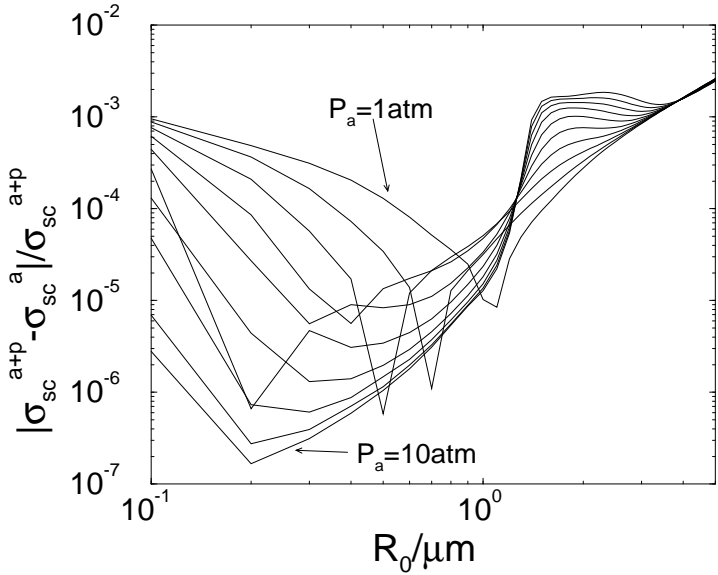,width=10cm,angle=-90}}
\end{picture}
\end{center}
\caption[]{Relative error of scattering cross section computations with
and without passive emission taken into account. The numerical calculations
were performed using the complete formula (\ref{sigmasc}) with or without
the passive contribution and driving with an ultrasound diagnostics pulse
according to (\ref{pulse}).
The curves correspond to
driving pressure amplitudes $P_a$ of $1-10$\,atm, in steps of 1\,atm.
The error is nowhere greater than about $0.25\%$.
}
\label{figerror}
\end{figure}

With $P(t)$ given by (\ref{pulse}), $I_{inc}$ can be computed, so that
with (\ref{ps}), (\ref{rp}) and (\ref{sigma}--\ref{wsc}),
$\sigma_{sc}$ results. Let us first
convince ourselves that the passive part of (\ref{ps}) is tiny
in the nonlinear case, too. In Fig.~\ref{figerror} we present
the {\em relative errors} made in determining scattering cross sections
from $P_s^a(r,t)$ (or $q_s^a(t)$) alone rather than from the full
interference formula (\ref{sigmasc}). It is easy to see that these errors
are nowhere greater than about $0.25\%$, and in most cases much smaller.
The huge advantage of (actively emitting) bubbles as scatterers compared
to stiff solid bodies of similar size is here once more strikingly
demonstrated.
For computational simplicity,
the cross sections presented in the following figures were computed using
only the active emission part. 

\begin{figure}[htb]
\setlength{\unitlength}{1.0cm}
\begin{center}
\begin{picture}(10,8)
\put(-1.,0.){\psfig{figure=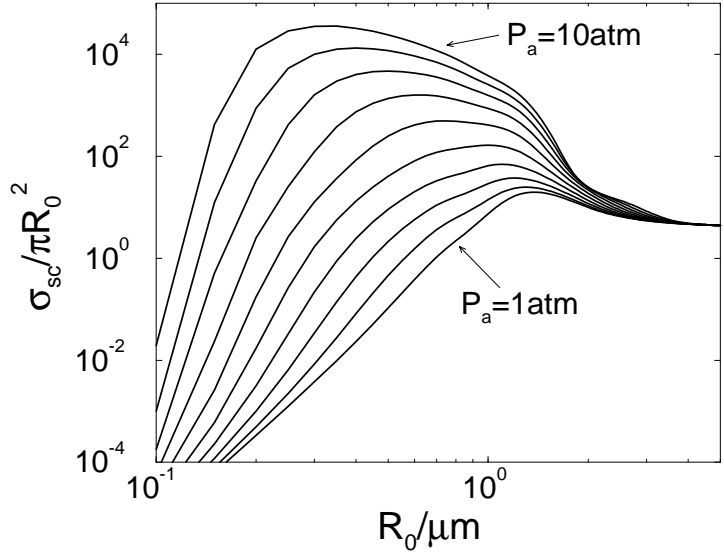,width=10cm,angle=-90}}
\end{picture}
\end{center}
\caption[]{Total scattering cross sections of bubbles driven by ultrasound
pulses. From bottom to top: $P_a=1-10$\,atm in steps of 1\,atm. The
linear profile of $\sigma_{sc}$ is drastically changed for high $P_a$,
the resonance radius shifts to smaller $R_0$ due to nonlinearities and
the cross sections are much larger around the resonance radius than in
the linear case (up to 3-4 orders of magnitude).
}
\label{fullscatt}
\end{figure}

Fig.~\ref{fullscatt} shows
the normalized cross sections for various $R_0\in\{0.1\mu m, 5.0\mu m\}$
and $P_a=1-10$atm (in 1\,atm intervals). These values represent
typical bubble
sizes found in diagnostic bubble suspensions (often the size distribution
shows a peak around $R_0\approx 1\mu m$ \cite{nan93}) and typical sound
pressure amplitudes in the focus of clinical ultrasound devices. From
linear analysis, we know what the cross sections for $P_a\to 0$ look like.
This curve is readily reproduced for small $P_a$.
While the curve for 1\,atm
is still essentially unaltered compared to the linear case, there are
dramatic changes in $\sigma_{sc}(R_0)$ with higher $P_a$. E.g., the
resonance radius $R_0^{res}$ undergoes a shift away from its linear
value to smaller $R_0$, i.e., in the opposite direction of the shift
induced by the polychromatic pulse. This shift is explained by the
nonlinearity of the oscillator (cf.\ \cite{ll60}):
when expanding the RP equation up to nonlinearities of third order,
one obtains for the nonlinear eigenfrequency
\be
\om_0^{NL}=\om_0+a^2\left({3\beta\over 8\om_0}-{5\alpha^2\over 12\om_0^3}
\right)\,,
\label{rpnl}
\ee
with the amplitude $a$ of the oscillation and
\be
\alpha=-\left(2\om_0^2-{2\zeta\over\rho_l R_0}\right), \quad
\beta={10\over 3}\om_0^2 +{14\zeta\over 3\rho_l R_0}\,.
\label{alphabet}
\ee
It is easy to see that with these values we always have $\om_0^{NL}<\om_0$,
i.e., the bubble is an oscillator with a {\em soft}
potential. With stronger nonlinearity, $\om_0^{NL}$ becomes smaller for a
given $R_0$; in
order to be in resonance, we require $\om_d=\om_0^{NL}$ with constant
$\om_d$. Therefore, $R_0$ must be {\em decreased} in order to increase
$\om_0$ beyond $\om_d$ to ensure that the resonance condition
$\om_d=\om_0^{NL}$ is again fulfilled. Thus, the nonlinear resonance
radius $R_0^{res,NL}$ is smaller than its linear counterpart.
Note also that the resonance structure is blurred especially for high
driving; this is in contrast to the case of monochromatic driving
\cite{gro97_zom},
where well-defined resonance radii can be recognized up to the highest $P_a$.

A most striking feature of Fig.~\ref{fullscatt} is the tremendously enlarged
scattering cross section, especially at
radii smaller than the linear resonance radius (i.e., in the region of nonlinear
resonance).
Sometimes, $\sigma_{sc}$ is greater by several orders of magnitude compared
to the linear case (note the logarithmic
scale). Only for $R_0\gg R_0^{res,lin}$ is the shape of the curve virtually
unaltered. Thus, the scattering cross sections for {\em small} bubbles
(below $\approx R_0=1\mu m$) are severely {\em underestimated}
 by the linear theory.

The reason for this effect is found in the typical $R(t)$ radius dynamics
of nonlinearly driven bubbles: when $P_a$ is sufficiently high, they undergo
violent {\em collapsing} instead of just a smooth oscillation. This leads
to high velocities and extraordinarily high accelerations of the bubble wall,
exceeding $10^9g$ in the solutions of (\ref{rp}) for the highest $P_a$ of
our study
(even larger accelerations are known in other contexts
of bubble dynamics, see \cite{bar97}).
These values of $\dot{R}$ and especially $\ddot{R}$,
which far exceed those expected for linear response, lead to very high active
emission pressures (cf.\ Eq.~(\ref{psa})) and thus to a huge $\sigma_{sc}$.

\begin{figure}[htb]
\setlength{\unitlength}{1.0cm}
\begin{center}
\begin{picture}(10,8)
\put(-1.,0.){\psfig{figure=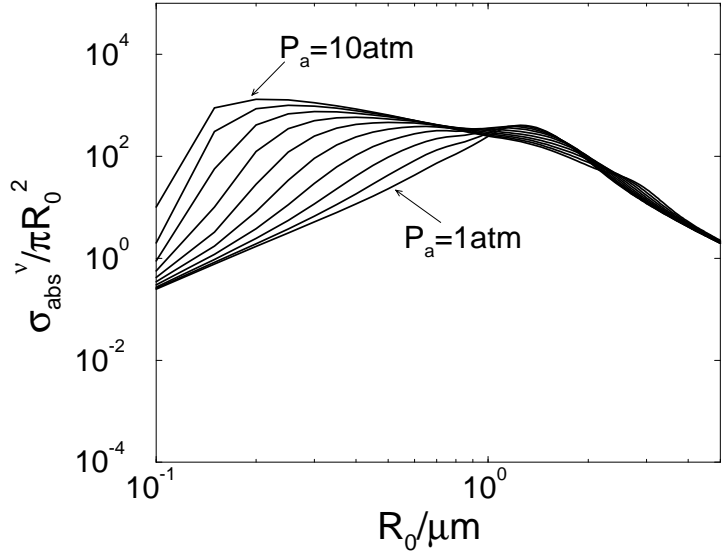,width=10cm,angle=-90}}
\end{picture}
\end{center}
\caption[]{Total viscous absorption cross sections of bubbles driven by
ultrasound pulses. From bottom to top: $P_a=1-10$\,atm in steps of 1\,atm. The
change in shape with growing $P_a$ is not as drastic as for the scattering
cross sections. $\sigma_{abs}^{\nu}$ is considerably smaller than
$\sigma_{sc}$ for the highest $P_a$, while it is much larger in the linear
case of small driving.
}
\label{fullabs}
\end{figure}

\subsection{Absorption cross sections}
In Fig.~\ref{fullabs} the viscous absorption cross section $\sigma_{abs}^{\nu}$
computed from (\ref{viscdiss}) and (\ref{sigmaabs}) is shown for the
same parameter combinations as for $\sigma_{sc}$ in Fig.~\ref{fullscatt}.
As expected, the resonance peak undergoes similar nonlinear shifting.
However, although $\sigma_{abs}^{\nu}$ also shows a tendency to grow for higher
$P_a$, it is obvious
that its dependence on driving pressure amplitude is nowhere near as
dramatic as in the case of the scattering cross section. This is because
in the computation of $W_{dis}^{\nu}$ via (\ref{viscdiss}), only the
velocity $\dot{R}$ is present, but not the acceleration, which is
responsible for the tremendous sound emission, as argued in the previous
section. As usual, the radiative processes are
connected with acceleration, while the dissipative processes are
governed by velocity. 
For high $P_a$
we can therefore conclude that -- regardless of the size of the bubble --
most of the incident sound energy is scattered again into sound, and not
dissipated via viscous forces.

\subsection{Secondary absorption}
\begin{figure}[htb]
\setlength{\unitlength}{1.0cm}
\begin{center}
\begin{picture}(10,8)
\put(-1.,0.){\psfig{figure=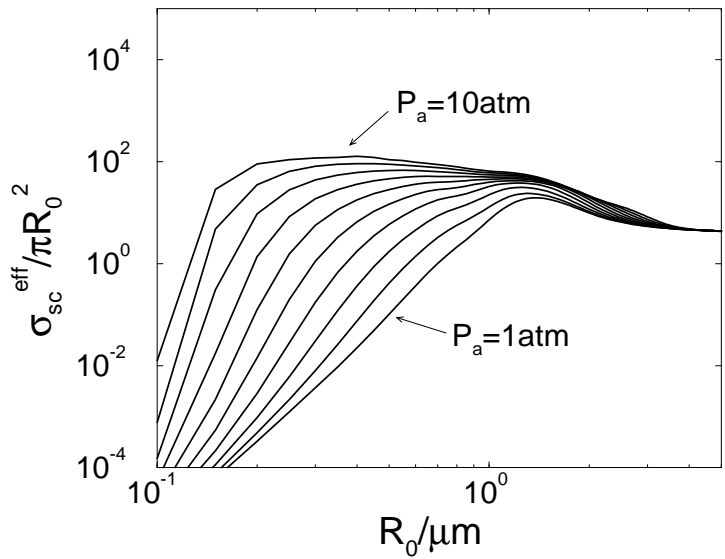,width=10cm,angle=-90}}
\end{picture}
\end{center}
\caption[]{Effective scattering cross sections of bubbles driven by
ultrasound pulses and separated from the detector by a water layer of
5cm width, which acts to damp out high frequency sound according to 
(\ref{waterabs}).
From bottom to top: $P_a=1-10$\,atm in steps of 1\,atm.
The growth for large $P_a$ is much diminished here as compared to
Fig.~\ref{fullscatt}. The high frequency components of the sound emitted
by the small bubbles at high $P_a$ are absorbed in water and thus the
major part of the emitted sound energy is lost to the detector.
}
\label{dampscatt}
\end{figure}

It is tempting to conclude that for the highest $P_a$ we have the optimal
situation for detection of scattered sound. This is not necessarily so, 
however, because
of the {\em spectrum} of the emitted acoustic radiation.
We show in a separate paper \cite{hil98}
that the most strongly driven bubbles emit
sound in a spectrum of immense band width, with a large portion of the
energy in the ultra high frequency part (GHz). These frequencies are
readily absorbed (on a length scale of $cm$ or less)
by water or other media (blood, tissue) frequently
encountered in diagnostic applications (cf.\ e.g.\ \cite{seh82,jon86}).
With this absorption, most of the
sound energy is converted into heat. Fig.~\ref{dampscatt} shows the
example of the
{\em effectively} detected scattered 
sound after the pulses have traveled through
a 5\,cm layer of water. The absorption properties of water are well known:
the energy contained in every Fourier component (frequency $f=\om/2\pi$) 
of the emitted sound signal decays over a distance $r$ like 
\begineq
E_s(r,f) = E_s(0,f)\exp(-\alpha_w f^2 r)\,,
\label{waterabs}
\endeq
with the known characteristic
absorption coefficient of water $\alpha_w\approx 1.5\cdot 10^{-14} s^2/m$ 
\cite{her59}. Thus, the highest frequency components experience the strongest
damping, and it is 
especially the high-power, high-frequency emission of
scatterers at small $R_0$ which is radically diminished by this 
process, although the cross sections do not drop to or near
their linear values. The implications for diagnostics, such as possible
risks of the heat deposition connected with the absorption, are dealt with
in separate work \cite{hil98}.

\section{Summary and conclusions}\label{summary}

We have presented a unified view of
gas bubble sound scattering and absorption.
Starting from the description of the bubble as a nonlinear oscillator,
general formulas
have been developed that can be shown to transform, in certain limiting cases,
into a number of apparently discrepant formulas found in literature, whether
they be derived from oscillator theory \cite{lei96}, scattering theory
\cite{nis75} or the theory of solid body sound scattering \cite{ll87}.
It is important to state that there is
{\em no antagonism} between ``scattering'' and ``active sound emission''.
Both terms refer to the sound that can be detected after incident sound has
hit a gas bubble. The total scattering cross section does contain a
contribution of {\em passive} sound emission, but this contribution is
tiny and can always be neglected for situations in ultrasound diagnostics.
Bubbles driven in the {\em nonlinear} range show
a {\em much larger} scattering cross section than in the linear case and are
more ``actively'' emitting sound in the sense that the incident energy is
primarily converted into sound and not into viscous heating, as is
the case for linear driving. The most prodigiously radiating
bubbles, at small
radii and large driving amplitude, emit such high frequency sound that it is
again absorbed by the surrounding medium and leads to secondary heating.
The effective yield of scattered sound energy is therefore much smaller than
what would be expected without taking absorption into account. Still,
the ultrasound scattering capabilities of bubbles and therefore their
effectiveness as a contrast agent in diagnostics are considerably enhanced
due to the nonlinearity of the oscillations.

\vspace{0.5cm}

\noindent
{\bf Acknowledgments:} Support by the DFG under the grants SFB\,185-D8 and
Lo 556/3-1 is gratefully acknowledged.


\begin{thebibliography}{10}

\vspace{-1.5cm}

\bibitem{nan93}
See the~articles in {\em Advances in echo imaging using contrast
  enhancement}, edited by N.~C. Nanda and R. Schlief (Kluwer Academic
  Publishers, Dordrecht, 1993).

\bibitem{gra68}
R. Gramiak and P.~M. Shah, Invest. Radiol. {\bf 3},  356  (1968).

\bibitem{ll87}
L.~D. Landau and E.~M. Lifshitz, {\em Fluid Mechanics} (Pergamon Press, Oxford,
  1987).

\bibitem{lei96}
T.~G. Leighton, {\em The acoustic bubble} (Academic Press, London, 1996).

\bibitem{nis75}
R.~Y. Nishi, Acustica {\bf 33},  65  (1975).

\bibitem{gro97_zom}
S. Grossmann, S. Hilgenfeldt, D. Lohse, and M. Zomack, J. Acoust. Soc. Am. {\bf
  102},  1223  (1997).

\bibitem{pro74}
A. Prosperetti, J. Acoust. Soc. Am {\bf 56},  878  (1976).

\bibitem{pro75}
A. Prosperetti, J. Acoust. Soc. Am {\bf 57},  810  (1975).

\bibitem{hil98}
S. Hilgenfeldt, D. Lohse, and M. Zomack, {\em Scattering and heat deposition 
of pulse-driven microbubbles}, preprint (1998).

\bibitem{bre95b}
C.~E. Brennen, {\em Cavitation and Bubble Dynamics} (Oxford University Press,
  Oxford, 1995).

\bibitem{ray17}
Lord Rayleigh, Philos. Mag. {\bf 34},  94  (1917).

\bibitem{gil52}
F.~R. Gilmore, Hydrodynamics Laboratory, California Institute of Technology,
  Pasadena, report {\bf 26-4},    (1952).

\bibitem{fly75}
H.~G. Flynn, J. Acoust. Soc. Am. {\bf 58},  1160  (1975).

\bibitem{kel80}
J.~B. Keller and M.J. Miksis, J. Acoust. Soc. Am. {\bf 68},  628  (1980).

\bibitem{las81}
G.~J. Lastman and R.~A. Wentzell, J. Acoust. Soc. Am. {\bf 69},  638  (1981).

\bibitem{las82}
G.~J. Lastman and R.~A. Wentzell, J. Acoust. Soc. Am. {\bf 71},  835  (1982).

\bibitem{pro91}
A. Prosperetti, J. Fluid Mech. {\bf 222},  587  (1991).

\bibitem{pro94}
A. Prosperetti,  in {\em Bubble dynamics and interface phenomena}, edited by
  J.~Blake et~al. (Kluwer Academic Publishers, Dordrecht, 1994), p.\ 3.

\bibitem{bar97}
B.~P. Barber {\it et~al.}, Phys. Rep. {\bf 281},  65  (1997).

\bibitem{lan69}
H. Landolt and R. B\"ornstein, {\em Zahlenwerte und Funktionen aus Physik und
  Chemie} (Springer, Berlin, 1969).

\bibitem{ple77}
M. Plesset and A. Prosperetti, Ann. Rev. Fluid Mech. {\bf 9},  145  (1977).

\bibitem{hil96}
S. Hilgenfeldt, D. Lohse, and M.~P. Brenner, Phys. Fluids {\bf 8},  2808
  (1996).

\bibitem{bje09}
V. Bjerknes, {\em Die Kraftfelder} (Friedrich Vieweg, Braunschweig, 1909).

\bibitem{pro84b}
A. Prosperetti, Ultrasonics {\bf 22},  115  (1984).

\bibitem{dej91}
N. deJong~et al., Ultrasonics {\bf 29},  324  (1991).

\bibitem{ll60}
L.~D. Landau and E.~M. Lifshitz, {\em Mechanics} (Pergamon Press, Oxford,
  1960).

\bibitem{seh82}
C.~M. Sehgal and J.~F. Greenleaf, J. Acoust. Soc. Am. {\bf 72},  1711  (1982).

\bibitem{jon86}
H.~A.~H. Jongen, J.~M. Thijssen, M. van~den Aarssen, and W.~A. Verhoef, J.
  Acoust. Soc. Am. {\bf 79},  535  (1986).

\bibitem{her59}
K.~F.~Herzfeld and T.~A.~Litovitz,
{\em Absorption and dispersion of ultrasonic waves} (Academic Press, 
New York, 1959).


\end{thebibliography}
\end{document}